# Hybrid acousto-electromagnetic metamaterial superconductors

Igor I. Smolyaninov [1,2)], Vera N. Smolyaninova [3)]

[1]*Department of Electrical and Computer Engineering, University of Maryland, College Park, MD 20742, USA*

[2] *Saltenna LLC, 1751 Pinnacle Drive #600 McLean, VA 22102, USA*

[3]*Department of Physics Astronomy and Geosciences, Towson University,*

*8000 York Rd., Towson, MD 21252, USA*

**Recent theoretical and experimental studies demonstrated that an electromagnetic metamaterial approach is capable of drastically increasing the critical temperature $T_c$ of composite metamaterial superconductors. This progress was achieved by engineering the frequency-dependent dielectric response function of the metamaterial. Here we further extend this approach by introducing the concept of hybrid acousto-electromagnetic hyperbolic metamaterial superconductors, which are projected to exhibit strongly enhanced superconducting properties due to artificially engineered broadband divergency of the Eliashberg electron-phonon spectral density function $\alpha^2 F(\omega)$. Based on the developed approach, a hybrid acousto-electromagnetic metamaterial is proposed, which is made of LSCO/STO nano-alloy.**

Recent theoretical [1,2] and experimental [3-5] studies demonstrated that an electromagnetic metamaterial approach is capable of drastically increasing the critical temperature of composite metamaterial superconductors. Metamaterials are made from assemblies of multiple elements arranged in repeating subwavelength structures (which are sometimes called "artificial atoms"), and they derive their macroscopic properties from the properties of these newly designed subwavelength structures. It appears that several approaches developed in the electromagnetic metamaterial field may be utilized to engineer artificial metamaterial superconductors, which demonstrate considerably improved superconducting properties. This initial progress was based on the well-known relationship between the superconducting properties of a material and its dielectric response function $\varepsilon_{eff}(q,\omega)$. Indeed, over fifty years ago Kirzhnits and co-workers [6] established that the critical temperature, $T_c$, of a superconductor may be related to the behaviour of $\varepsilon_{eff}^{-1}(q,\omega)$ near its poles. This macroscopic electrodynamics framework describes electron-electron interaction in a superconductor using an effective Coulomb potential

$$V(\vec{q},\omega) = \frac{4\pi e^2}{q^2 \varepsilon_{eff}(\vec{q},\omega)} = \frac{V_C}{\varepsilon_{eff}(\vec{q},\omega)}, \qquad (1)$$

where $\varepsilon_{eff}(q,\omega)$ is the dielectric response function of the superconductor, which is treated as an effective medium, and $V_C = 4\pi e^2/q^2$ is the Fourier-transformed Coulomb potential in vacuum. Therefore, the recently developed ability to engineer a desired dielectric response function $\varepsilon_{eff}(q,\omega)$ in a composite electromagnetic metamaterial gives us an ability to increase its $T_c$ [1]. For example, considerable enhancement of the attractive electron-electron interaction in epsilon near zero (ENZ) [7] and hyperbolic metamaterials [8] was related to the increase of $T_c$ in Al/Al$_2$O$_3$-based ENZ [3] and hyperbolic [4] metamaterial structures, respectively. In both cases $\varepsilon_{eff}^{-1}(q,\omega)$ of the metamaterial exhibits additional poles compared to the bulk aluminium. These additional poles were identified as plasmon-phonons of the composite metamaterial structures, and they were observed using inelastic neutron scattering [5]. The hybrid acousto-electromagnetic nature of these excitations gives us a strong hint that further improvement of the superconducting properties of artificial metamaterials should arise from combined use of both acoustic metamaterial and electromagnetic metamaterial toolboxes.

In this paper we further extend the metamaterial superconductor approach by introducing the concept of hybrid acousto-electromagnetic metamaterial superconductors. Implementation of the acoustic metamaterial tools seems to be a natural extension of the earlier purely electromagnetic approach. Using the acoustic metamaterial approach, we should be able to engineer the phonon density of states with the goal of increased $T_c$ in a superconductor. Some elements of this approach were already implemented in [9], where it was proposed to alter the phonon density of states in a thin superconducting film by perforating the film with a periodic nanohole array. Here we report a much more comprehensive approach, which is aimed at re-engineering of the Eliashberg electron-phonon spectral density function $\alpha^2 F(\omega)$ [10] in a hybrid acousto-electromagnetic metamaterial, where $F(\omega)$ is the phonon density of states, and $\alpha(\omega)$ is the electron-phonon coupling constant at a given frequency $\omega$.

Within the scope of Eliashberg theory the superconducting transition temperature $T_c$ may be approximated [9] as

$$k_B T_C = \frac{\hbar \omega_{\ln}}{1.2} \exp\left(-\frac{1.04(1+\lambda)}{\lambda - \mu^*(1+0.62\lambda)}\right), \qquad (2)$$

where $\mu^*$ is the Coulomb pseudopotential, $\lambda$ is the electron-phonon coupling constant, $\omega_{ln}$ is the logarithmic average of the phonon frequencies defined by

$$\omega_{\ln} = \exp\left(\frac{2}{\lambda}\int_0^\infty d\omega \ln(\omega)\frac{\alpha^2 F(\omega)}{\omega}\right), \quad (3)$$

and $\alpha^2 F(\omega)$ is the Eliashberg spectral function. Under several approximations [11], the Coulomb pseudopotential is given by

$$\mu^* = \frac{\mu}{1+\mu\ln(E_F/\omega_c)}, \quad (4)$$

where $E_F$ is the Fermi energy, $\omega_c$ is a cut-off frequency, which is usually chosen as ten times the maximum photon frequency, and $\mu$ is an average electron-electron matrix element times the density of states at the Fermi level.

The Eliashberg spectral function $\alpha^2 F(\omega)$ is the key quantity of the theory, which is defined as

$$\alpha^2 F(\omega) = \frac{1}{N(0)}\sum_{k,k',i,j,\nu}\left|g_{k,k'}^{ij,\nu}\right|^2 \delta(\varepsilon_k^i)\delta(\varepsilon_{k'}^j)\delta(\omega-\omega_{k-k'}^\nu), \quad (5)$$

where $N(0)$ is the single-spin electron density of states at the Fermi level, and $g_{k,k'}^{ij,\nu}$ is the electron-phonon matrix element, with $\nu$ being the phonon polarization index, and $k$, $k'$ representing electron wave vectors with band indices $i$ and $j$, respectively. $\varepsilon_k^i$ denotes the band energy of an electron measured with respect to the Fermi energy. If the Eliashberg function is known, then a simple integration gives the total electron-phonon coupling constant

$$\lambda = 2\int_0^\infty \frac{d\omega}{\omega}\alpha^2 F(\omega) \quad (6)$$

It was demonstrated by analytical and numerical solutions of the Eliashberg equations that the spectral shape of $a^2F(\omega)$ plays a significant role in defining the ultimate $T_c$ of a superconductor [12]. It appears that $T_c$ decreases with an increase of the spectral width of $a^2F(\omega)$, while the highest $T_c$ is achieved when $a^2F(\omega)$ exhibits an Einstein spectrum (close to a δ-function at some phonon frequency $\omega_0$). Below we will demonstrate that such a spectrum may be engineered using an acoustic hyperbolic metamaterial approach.

Following the development of electromagnetic hyperbolic metamaterials, which exhibit broadband divergence in their photonic density of states [8], somewhat related designs of acoustic hyperbolic metamaterials were introduced [13], which exhibit similar divergence in their phonon density of states (DOS). Hyperbolic metamaterials are one of the most important classes of metamaterials which exhibit extremely anisotropic behaviour due to anisotropic design of their artificially engineered subwavelength constitutive elements. In the case of acoustic hyperbolic metamaterials they are designed to have extremely anisotropic effective densities $\rho(\omega)$. In a two dimensional metamaterial the dispersion relation of phonons in the metamaterial is given by

$$\frac{k_x^2}{\rho_x} + \frac{k_y^2}{\rho_y} = \frac{\omega^2}{B}, \qquad (7)$$

where $k$ and $\omega$ are the phonon wave number and frequency, respectively, and $B$ is the bulk modulus of the medium [13]. If both $\rho_x$ and $\rho_y$ are positive, the equifrequency contour of the phonon dispersion law is an ellipse, and the phonon DOS is finite. On the other hand, if $\rho_x$ and $\rho_y$ have opposite signs, this equifrequency contour becomes a hyperbola. In such a case, in the continuous medium approximation the phonon DOS of the metamaterial becomes nominally infinite in the hyperbolic frequency bands (see Fig.1a). As discussed above, based on the Eliashberg theory, such a case would be beneficial for $T_c$ increase of a metamaterial superconductor. Note however that the total number of degrees of freedom associated with atomic movements is defined by the total number of atoms in the system, and therefore it is fixed. As a result, creation of a hyperbolic band inside an acoustic metamaterial just redistributes the spectral weight of phonon states between different frequency bands, without changing the total number of phonon states.

A typical example of a hyperbolic acoustic metamaterial geometry is presented in Fig.1b. This metamaterial geometry is made of periodic linear chains of acoustic resonators arranged along the $x$ direction inside a background matrix material. The acoustic resonators are separated by periodic gaps in the orthogonal $y$ direction. While $\rho_x$ of such an acoustic metamaterial remains positive and almost frequency independent, $\rho_y$ changes sign near the acoustic resonator eigenfrequency. Indeed, as was demonstrated in [13], phonon propagation in x- and y-directions can be decoupled in this type of structure. Propagation in the x-direction is close to phonon propagation in the unperturbed background matrix material (since there is no periodic structure in this direction), and the effective density $\rho_x$ is close to the background material density. On the other hand, the frequency dependent $\rho_y$ may be calculated using a lump element-based acoustic circuit model [14], which gives

$$\rho_y = \frac{Z_a}{i\omega} \frac{1}{da^2}, \qquad (8)$$

where the acoustic impedance $Z_a$ is obtained as

$$Z_a = i\omega m_a + \frac{1}{i\omega C_a} \quad . \tag{9}$$

In the latter equation $m_a$ and $C_a$ denote the acoustic mass and the acoustic capacitance of the resonator, respectively [14]. Near the resonant frequency, which is defined by Eq.(9), $\rho_y$ changes sign, and the acoustic metamaterial becomes hyperbolic.

While the hyperbolic acoustic metamaterial geometry presented in Fig.1b is highly idealized, it may be reasonably approximated in a realistic metamaterial superconductor. One such possibility would be to nanofabricate an appropriate pattern of grooves in a thin superconducting film. Another possibility is to use thin superconducting films grown by nano-alloying [15,16], in which different material phases may self-separate into nano-columnar structures. In order for the metamaterial description of such a structure to remain valid, the wavelength of phonons in the hyperbolic band $\lambda_h$ must be considerably larger than the geometric parameters of the structure (such as the average distance between the nano-columns and their diameter). Let us demonstrate that this condition may be met in a typical high $T_c$ superconductor.

An example of a potentially interesting metamaterial superconductor structure is presented in Fig.2, which is based on the reported study of strain-induced superconductivity in $La_2CuO_{4+\delta}$ using a simple vertically aligned nanocomposite approach [15]. It is made of strontium titanate (STO) nanocolumns inside a superconducting $La_{1.85}Sr_{0.15}CuO_4$ (LSCO) matrix. The STO nanocolumns in this structure play the role of acoustic resonators shown in Fig.1b. Within a parameter range estimated below, this structure will behave as a hyperbolic acoustic metamaterial. Moreover, it may also behave as an electromagnetic ENZ metamaterial within an overlapping parameter range, which results in this metamaterial structure acting as a hybrid acousto-electromagnetic metamaterial. The proposed design targets the low frequency optical phonon band in LSCO, which is located around $\nu = 3.6$ THz (see Fig.3a). It has been established that these optical phonon modes have a large density of states $F(\omega)$, which leads to considerable increase in the superconducting coupling constant $\lambda$ [17]. Moreover, similar low frequency optical phonon modes also exist in other high $T_c$ cuprates. Further increase of the phonon DOS in this frequency range via metamaterial engineering is expected to increase the critical temperature of the metamaterial. Such a hybrid metamaterial superconductor would potentially exhibit much higher $T_c$ compared to the parent LSCO superconductor because of the Eliashberg electron-phonon spectral density function $a^2F(\omega)$ exhibiting an Einstein spectrum, and because of the effective Coulomb potential (1) exhibiting ENZ behaviour. We should also note that the described effect may also provide an additional contributing factor in the $T_c$ increase of $La_2CuO_{4+\delta}$, which was observed in [15] and which was attributed to strain effects.

The key observation making the acoustic metamaterial approach possible is that optical phonons shown schematically in Fig.2b may have wavelengths that are sufficiently large compared to the geometrical parameter $d$ describing the metamaterial, and compared to the lattice periodicity $a$ of LSCO. Unlike very low frequency long-wavelength acoustic phonons, which are generally believed to play almost no role in the Cooper pairing of superconducting electrons, long wavelength optical phonons may play very important role in the electron pairing, and they can be affected by the metamaterial nanostructuring. Based on the frequency of these phonons and the velocity of sound in STO ($V_{STO}$ ~7900 m/s [18]), we may also estimate typical dimensions of the STO nanocolumns shown in Fig.2a. These dimensions must be of the order of

$$d \sim \lambda_{STO} \sim V_{STO}/\nu \sim 2.2 \text{ nm} \tag{10}$$

for the STO nanocolumn to behave as an acoustic resonator. This estimate indicates that STO nanocolumnar acoustic resonators may indeed be fabricated inside an LSCO film using nano-alloying [15, 16]. We should also note that due to its much higher velocity of sound ($V\sim 12000$ m/s) diamond may be even better choice for an acoustic nano-resonator material. However, diamond is much less amenable to standard nanofabrication techniques.

Let us now consider the electromagnetic properties of this metamaterial design in the 3.6 THz range. As illustrated in Fig.3, which compares phonon DOS of LSCO and STO based on results reported in [19,20], the density of phonon states of STO in the 3.6 THz range is low. On the other hand, the optical phonon modes of LSCO have a large density of states $F(\omega)$ in this frequency range, which is assumed to lead to considerable increase in the superconducting coupling constant $\lambda$ [17]. Following [21], a simplified dielectric response function of LSCO may be written as

$$\varepsilon_m(q,\omega) = \left(1 - \frac{\omega_p^2}{\omega^2 + i\omega\Gamma - \omega_p^2 q^2/k^2}\right)\left(1 - \frac{\Omega_1^2(q)}{\omega^2 + i\omega\Gamma_1}\right)\cdots\left(1 - \frac{\Omega_n^2(q)}{\omega^2 + i\omega\Gamma_n}\right) \tag{11}$$

where $\omega_p$ is the plasma frequency, $k$ is the inverse Thomas-Fermi radius, $\Omega_n(q)$ are dispersion laws of various phonon modes, and $\Gamma_n$ are the corresponding damping rates. Zeroes of the dielectric response function (which correspond to various phonon modes of LSCO) maximize electron-electron pairing interaction given by Eq. (1). By crossing these zeros as a function of frequency, the dielectric response function of LSCO changes sign. This argument and the phonon DOS data shown in Fig 3 indicate that the dielectric response functions of LSCO and STO must have opposite signs in the considerable portion of the 3.6 THz band. As a result, based on the theoretical consideration in [1-5], the metamaterial structure shown in Fig.2a may exhibit ENZ properties and additional plasmon-phonon modes. As demonstrated in [5], the plasmon-phonon modes manifest themselves as an additional pole of the inverse dielectric response function of the metamaterial

$$\varepsilon_{eff}^{-1} = \frac{n}{(3-2n)}\frac{1}{\varepsilon_m} + \frac{9(1-n)}{2n(3-2n)}\frac{1}{(\varepsilon_m + (3-2n)\varepsilon_d/2n)} \quad , \qquad . \tag{12}$$

where $n$ is the volume fraction of LSCO, and $\varepsilon_d$ is the dielectric response function of STO. Appearance of these hybrid plasmon-phonon modes (corresponding to the second pole in Eq. (12)) leads to an increase of the electron-phonon coupling constant $\alpha(\omega)$ of the metamaterial.

Using Eqs. (8, 9, 12) and based on the experimental data for density ($\rho_{LSCO}$ = 6.825 g/cm$^3$, $\rho_{STO}$ = 5.11 g/cm$^3$) and electromagnetic properties [22, 23] of LSCO and STO, respectively, we may estimate the geometrical parameters of a LSCO/STO metamaterial shown in Fig.2a which would simultaneously behave as a hyperbolic acoustic metamaterial and as an ENZ electromagnetic metamaterial in the 3.6 THz frequency range. Numerical simulations of acoustic and electromagnetic properties of such a metamaterial are illustrated in Fig.4. In these simulations it is assumed that the metamaterial sample thickness equals 3.3 nm, which produces N = 3 longitudinal acoustic resonance at 3.6 THz in the STO nanopillar. Based on [22], the plasma frequency of the LSCO was assumed to be $\omega_p$ = 290 THz. The volume fraction of LSCO in the metamaterial was assumed to be $n$ = 0.7.

Let us estimate a potential increase of the metamaterial $T_c$ due to re-engineered $\alpha^2 F(\omega)$ in the 3.6 THz range. Based on the consideration of the effect of the spectral shape of $\alpha^2 F(\omega)$ on $T_c$ reported in [12], it may be estimated as

$$T_C(R) = T_C(0)\exp\left(-\frac{\lambda^2}{(\lambda-\mu^*)^2}R\right), \tag{13}$$

where the spectral width $R$ of the Eliashberg function is defined as

$$R = \frac{1}{3} \frac{\langle (\omega - \langle \omega \rangle)^2 \rangle}{\langle \omega \rangle^2} \quad , \qquad (14)$$

and $T_c(0)$ is the hypothetic critical temperature of a superconductor assuming that its Eliashberg spectral function $\alpha^2 F(\omega)$ exhibits an Einstein δ-function-like spectrum. Based on Eqs. (13,14), a decrease of the spectral width $R$ compared to its starting magnitude in the bulk LSCO (defined by the phonon DOS depicted in Fig.3a) should lead to increased $T_c$ of the metamaterial superconductor compared to $T_c \sim 40$ K in bulk LSCO. Using the theoretically estimated value of $\lambda = 0.4$ [19] and $\mu^* \sim 0.1$- $0.2$ [24] for the optimally doped LSCO, and the spectral data in Fig.3a which give the spectral width of the order of $R \sim 0.15$, we have performed numerical modelling of the metamaterial $T_c$ as a function of R, which is shown in Fig.5. Based on this modelling, we conclude that the critical temperature of the LSCO/STO metamaterial may potentially double compared to bulk LSCO in the limit of δ-function-like Einstein phonon spectrum due to hyperbolic acoustic band formed around 3.6 THz. We should also note that a similar effect may also provide an additional contributing factor to the tripling of $T_c$ in $La_2CuO_{4+\delta}$ -based nanoalloys which was observed in [15] and which was attributed to strain effects.

In conclusion, in this work we have introduced a concept of hybrid acousto-electromagnetic hyperbolic metamaterial superconductors, which are projected to exhibit strongly enhanced superconducting properties due to artificially engineered broadband divergency of the Eliashberg electron-phonon spectral density function $\alpha^2 F(\omega)$. The proposed approach combines the acoustic and electromagnetic metamaterial tools to optimize $T_C$ of the composite metamaterial structure. Based on the developed approach, a hybrid acousto-electromagnetic metamaterial is proposed, which is made of LSCO/STO nano-alloy, and which is projected to exhibit superconducting critical temperature in the 80 K range.

This work was supported in part by ONR Awards N0001418-1-2681 and N00014-18-1-2653.

**Figure Captions**

**Figure 1**. (a) Comparison of an elliptic and a hyperbolic phonon dispersion laws in an acoustic metamaterial. In the case of a hyperbolic phonon dispersion law the phonon DOS becomes nominally infinite in the continuous medium approximation since the magnitude of phonon $k$ vector at a given frequency $\omega$ is unlimited. (b) A typical example of a hyperbolic acoustic metamaterial geometry. This metamaterial geometry is made of periodic linear chains of acoustic resonators (shown in yellow) arranged along the $x$ direction inside a background matrix material. The acoustic resonators are separated by periodic gaps in the orthogonal $y$ direction. While $\rho_x$ of such an acoustic metamaterial remains positive and almost frequency independent, $\rho_y$ changes sign near the acoustic resonator eigenfrequency.

**Figure 2**. (a) Realistic geometry of a hybrid acousto-electromagnetic metamaterial superconductor, which is made of STO nanocolumns inside a superconducting $La_{1.85}Sr_{0.15}CuO_4$ matrix. The STO nanocolumns in this structure play the role of acoustic resonators shown in Fig.1b. (b) Schematic representation of the acoustic and optical phonon modes. The acoustic metamaterial description is applicable in a shaded area where the phonon wavelength $\lambda$ remains considerably larger than the typical geometrical parameters $d$ describing the metamaterial, and the lattice periodicity $a$.

**Figure 3**. Comparison of the phonon DOS in LSCO and STO. (a) Theoretically calculated phonon DOS in optimally doped LSCO (from [19]). The low frequency optical phonon band in LSCO is located around 3.6 THz. (b) Experimentally measured Raman spectrum of STO (from [20]) indicates low phonon DOS in the 3.6 THz range (weak spectral lines marked by letter R correspond to the substrate).

**Figure 4**. (a) Calculated values of $\rho_x$ and $\rho_y$ of the LSCO/STO metamaterial superconductor as a function of phonon frequency. The metamaterial exhibits

hyperbolic behaviour below 3.6 THz. (b) Calculated dielectric response function of the LSCO/STO metamaterial superconductor. The metamaterial exhibits ENZ behaviour around 3.4 THz.

**Figure 5**. Simulated behaviour of $T_c$ of a LSCO/STO metamaterial as a function of spectral width $R$ of the Eliashberg function.

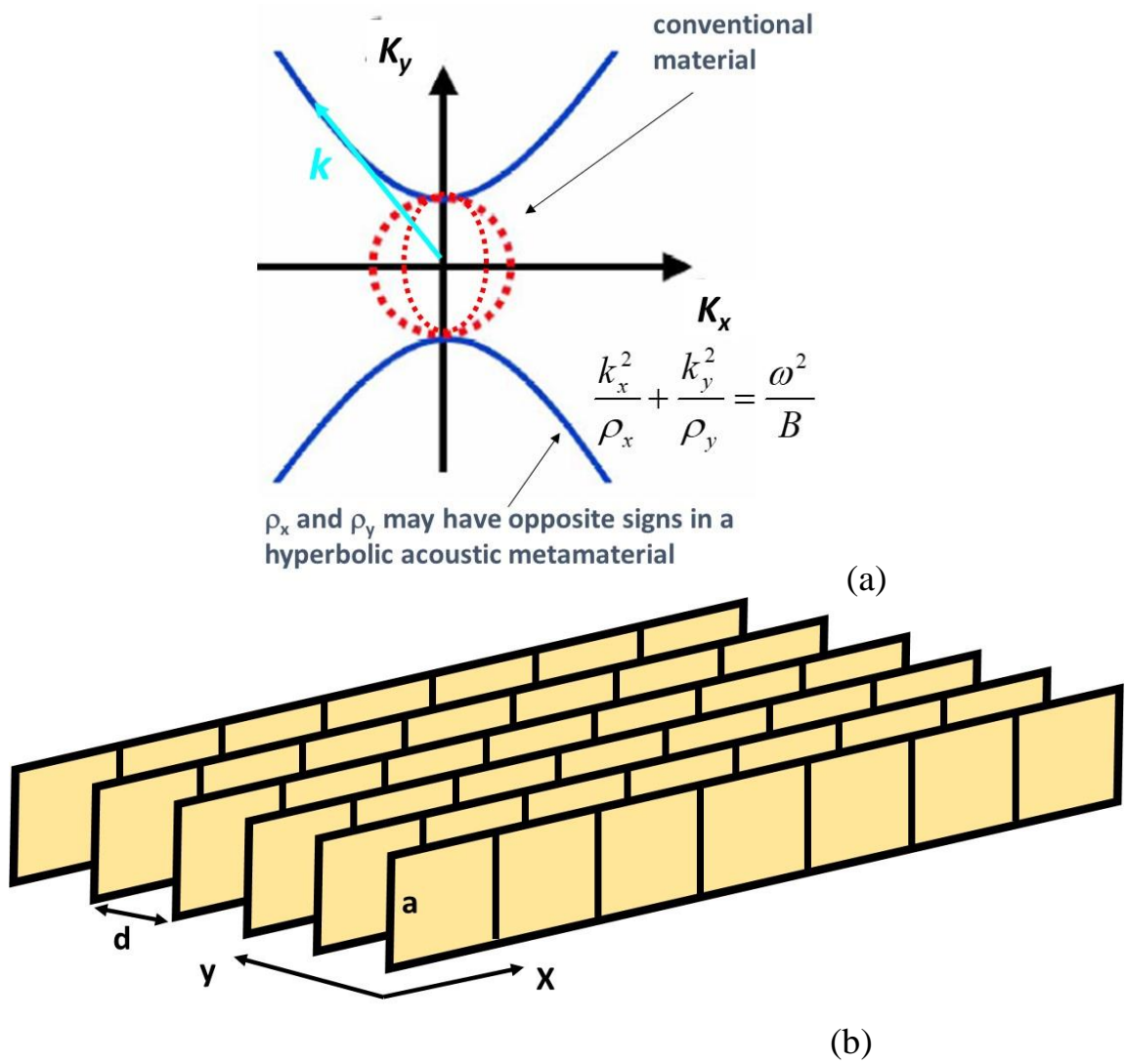

(a)

(b)

Fig. 1

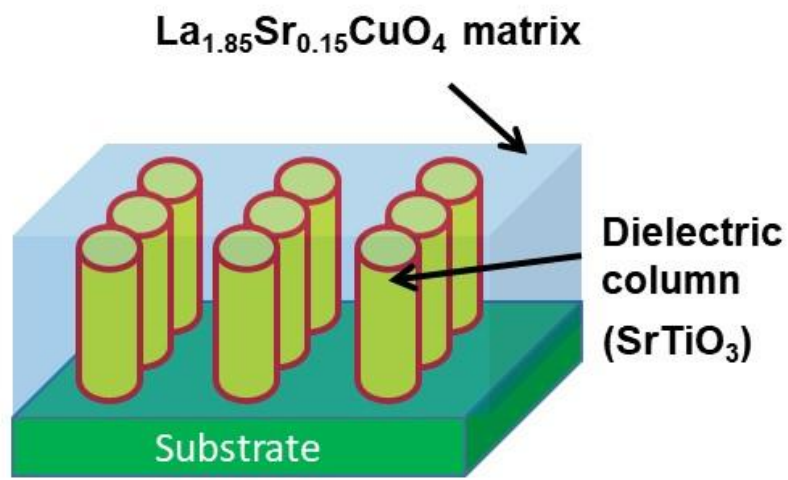

(a)

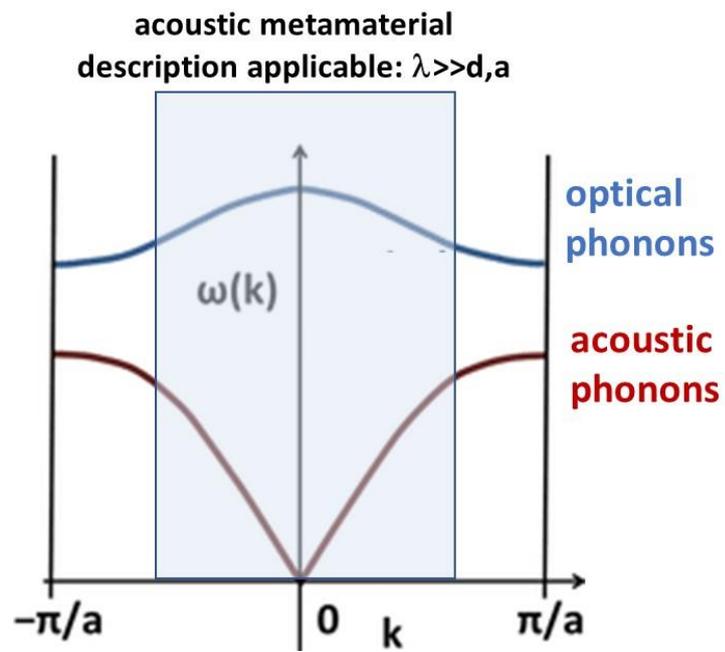

(b)

Fig. 2

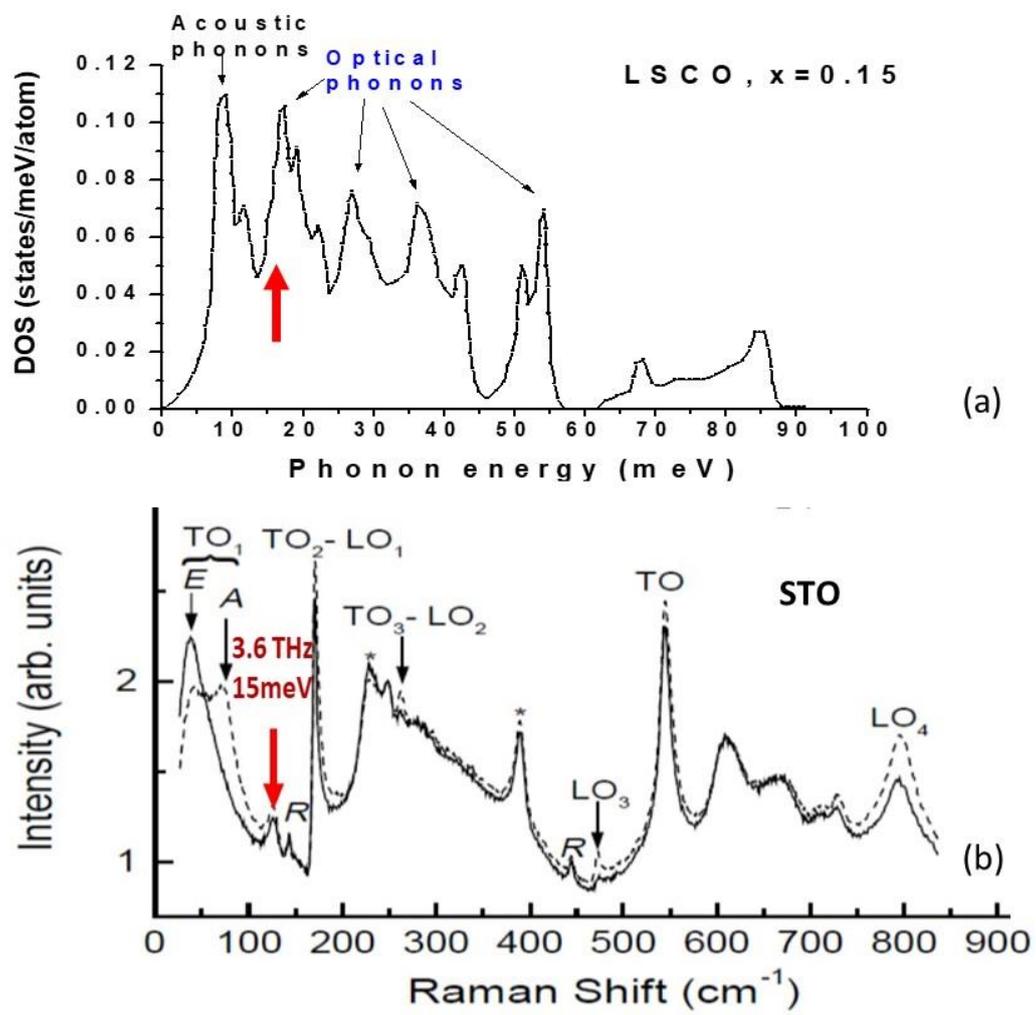

Fig. 3

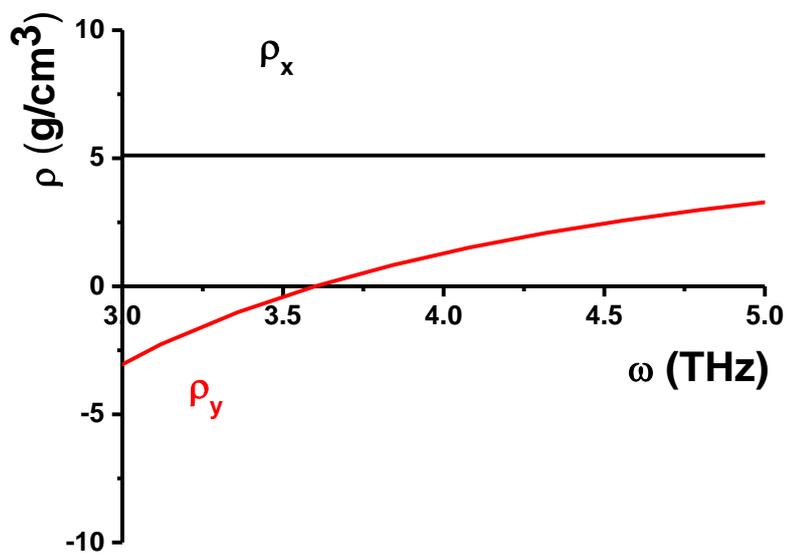

(a)

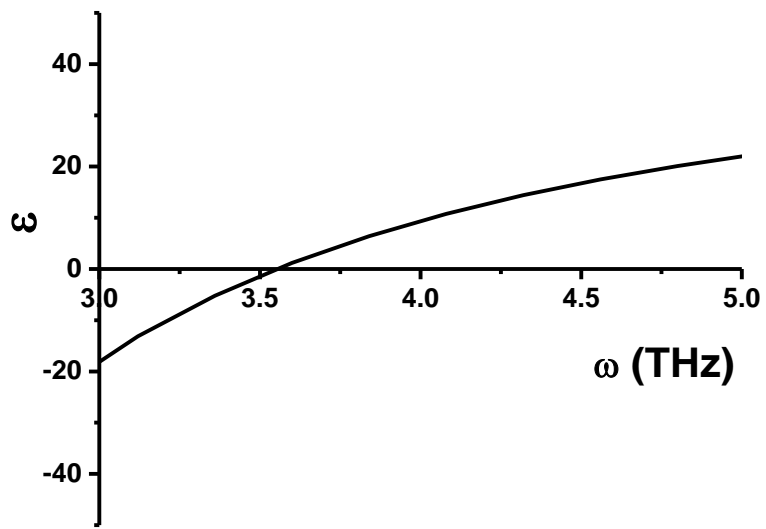

(b)

Fig. 4

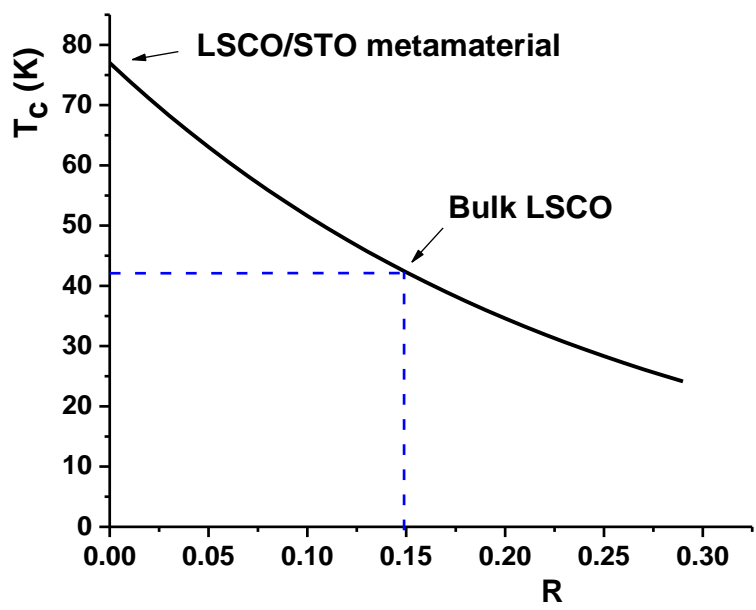

Fig. 5